\documentclass[12pt]{article}
\usepackage{amsmath}
\usepackage{latexsym}
\usepackage[dvips]{graphicx}
\usepackage{epsfig}
\usepackage[hang]{caption2}

\begin{document}
\begin{center}
{\Large An Adaptive Method for Valuing an Option on Assets with
Uncertainty in Stochastic Volatility }

\bigskip
Sergei Fedotov \footnote{%
Corresponding author. \newline \indent \emph{E-mail address:}
sergei.fedotov@manchester.ac.uk}, Stephanos Panayides \vskip 1.0cm
The School of Mathematics, The University of Manchester \\[0pt]
P.O Box 88, Manchester, M60 1QD, UK

\bigskip
The 5th of January 2006
\bigskip

\textbf{Abstract}

\end{center}
We present an adaptive approach for valuing the European call option
on assets with stochastic volatility. The essential feature of the
method is a reduction of uncertainty in latent volatility due to a
Bayesian learning procedure. Starting from a discrete-time
stochastic volatility model, we derive a recurrence equation for the
variance of the innovation term in latent volatility equation. This
equation describes a reduction of uncertainty in volatility which is
crucial for option pricing. To implement the idea of adaptive
control, we use the risk-minimization procedure involving random
volatility with uncertainty. By using stochastic dynamic programming
and a Bayesian approach, we derive a recurrence equation for the
risk inherent in writing the option. This equation allows us to find
the fair price of the European call option. We illustrate
numerically that the adaptive procedure leads to a decrease in
option price. \noindent
\begin{description}
\item[\textit{Keywords}:]  Stochastic Volatility, Adaptive Decision Process,
Bellman's equation.
\end{description}
\pagebreak
\section{Introduction}
Empirical observations on derivative prices show that implied
volatilities vary with strike price giving the well known volatility
``smile'' effect \cite{RH}. This suggests that the behavior of the
asset price, on which the option is written, may be captured by
models that recognize the stochastic nature of volatility (SV). One
can describe the underlying stock by a random process that is driven
by a random volatility (see, for example, \cite {Sircar,Lewis,Shep}
and references therein). A common feature of these models is that
the random volatility is described by a random process with
\textit{known} statistical characteristics. The problem with this
approach is that the volatility is not a tradeable asset, and
therefore the classical hedging can not be applied. As a result the
equation for an option price involves an unknown parameter, the
market price of volatility risk \cite{Sircar}. The difficultly is
that this parameter is not directly observable and one has to make
additional assumptions on the pricing of volatility risk. Moreover,
in practice it is very difficult to suggest the \textit{exact
}statistics for the volatility in advance. A different approach to
stochastic volatility has been suggested in \cite{Sircar2,Fed2}.
Instead of finding the exact option price one can focus on the
pricing bands for options. The other problem of stochastic
volatility models is associated with the efficient estimation of the
\textit{unobserved} volatility process from financial data. These
lead some researchers to accept the idea of \textit{uncertain}
volatility when all prices for the option are possible within some
range \cite{Av}. The question arises whether this uncertainty can be
reduced during decision making. One of the main purposes of this
work is to answer this question by using the simplest SV-model, the
idea of Bayesian learning procedure and adaptive decision process
(see \cite{Bellman,Press}). We suppose that some of the statistical
properties of volatility are not known initially. Instead, we assume
that we have an \textit{a priori} estimation for them. By using the
Bayesian
approach (see \cite{Press}), we revise these \textit{a priori }%
characteristics of random volatility. To implement the idea of
adaptive feedback control for option pricing, we use a
risk-minimization procedure (see \cite{Bu} and references therein)
and stochastic dynamic programming \cite{Bellman,Ber,J}. This work
extends the idea of using adaptive processes in option pricing
suggested in \cite{Fed}. Application of stochastic dynamic
programming for pricing of derivatives can be found in
\cite{Barron,Karoui}. It should be noted that the Bayesian learning
approach to option pricing was also used in
\cite{ACF,Bunnin,Gu,wolf} in different contexts. Bayesian estimation
of a stochastic volatility model by using option price was proposed
in \cite{Forbes}. The outline of the paper is as follows. In Section
2, we introduce a discrete-time stochastic volatility model and
describe the Bayesian learning procedure. We derive the recurrence
equation for the variance of the innovation term in latent
volatility equation. In Section 3, we describe the risk-minimization
procedure and derive the Bellman's equation for the risk inherent in
writing the option. By using this equation we find the fair price of
European call option. We illustrate numerically that the adaptation
procedure leads to a decrease in the option price.
\section{Uncertain Stochastic Volatility and Adaptation}
In this paper we consider a simple market with two traded assets: a
riskless bond,\ $B_{n},$\ and a risky asset (stock), $S_{n},$
evolving at discrete times $n=0,1,......N.$ The bond price $B_{n}$
is governed by the recurrence relation
\begin{equation}
B_{n+1}=(1+r)B_{n},\text{ \ \ \ \ \ \ }B_{0}>0\text{\ },
\label{bon}
\end{equation}
with the constant interest rate $r>0.$ The stock price $S_{n}$ is
governed by the stochastic difference equation
\begin{equation}
\ln \left( \frac{S_{n+1}}{S_{n}}\right) =\xi _{n},\text{ \ \ \ \ \ \ }%
S_{0}>0,  \label{basic}
\end{equation}
where the stochastic return $\xi _{n}$ is modelled as follows
\begin{equation}
\xi _{n}=\mu +\sigma \delta _{n}e^{h_{n}/2}.\ \   \label{aspr}
\end{equation}
Here $\mu $ is the mean return from holding a stock at time $n$,
$\sigma $ is the instantaneous volatility, $h_{n}$ is the
log-volatility (latent volatility) at time $n$ that follows a
stationary AR(1)-process:
\begin{equation}
h_{n+1}=\varphi h_{n}+u_{n},  \label{vol}
\end{equation}
where $\varphi $ is the``persistence'' parameter of volatility:
$|\varphi |<1 $. This is the simplest version of a stochastic
volatility model and gives a discrete-time approximation for
standard continuous stochastic volatility models (see, for example,
\cite{Shep,T}). There are two sources of uncertainty in stochastic
equations (\ref{aspr}) and (\ref{vol}), namely the innovation terms
$\delta _{n}$ and $u_{n}$. We assume that $\delta _{n}$ and $u_{n}$
are both Gaussian sequences of mutually independent random
variables, and $\delta _{n}$ has the following probability density
function:
\begin{equation}
\varphi (\delta )=\frac{d}{d\delta }\mathbf{P}\left\{ \delta
_{n}<\delta \right\} =\frac{1}{\sqrt{2\pi }}\exp \left\{
-\frac{\delta ^{2}}{2}\right\} . \label{ff}
\end{equation}
Let us now discuss the statistical properties of $u_{n}.$ The
empirical observations suggest that the log-volatility can be
reasonably approximated by the Gaussian distribution
\cite{An1,An2,Poon}. Suppose that the investor does not know the
exact value of the variance of $u_{n}.$ With enough information from
the past history, the investor is assumed to have an \textit{a
priori} value for it, such that the probability density function for
the first term, $u_{0},$ is \textit{\ \ }
\begin{equation}
p_{0}(u)=\frac{d}{du}\mathbf{P}\left\{ u_{0}<u\right\}
=\frac{1}{\sqrt{2\pi \sigma _{0}^{2}}}\exp \left\{
-\frac{u^{2}}{2\sigma _{0}^{2}}\right\} . \label{pp}
\end{equation}
Our idea is to use an adaptive procedure by which the uncertainty regarding $%
u_{n}$ can be reduced by using the equations
(\ref{basic})-(\ref{vol}). For this purpose one needs an equation
involving both random sequences, $u_{n}$ and $\delta _{n}$ in a
linear combination. From (\ref{basic})-(\ref{vol}) one can find
\begin{equation}
\ln S_{n+1}+h_{n+1}=\ln S_{n}+\mu +\sigma \delta
_{n}e^{h_{n}/2}+\varphi h_{n}+u_{n}.  \label{one}
\end{equation}
If we start with the given values of $S_{0}$ and $h_{0},$ it follows from (%
\ref{ff}) and (\ref{one}) \ that the likelihood of $\ln S_{1}+h_{1}$
conditional on $u_{0}=u$ is
\begin{equation}
L\left( \ln S_{1}+h_{1}|_{u}\right) =C_{L}\exp \left\{ -\frac{(\ln (\frac{%
S_{1}}{S_{0}})+h_{1}-\mu -\varphi h_{0}-u)^{2}}{\sigma ^{2}e^{h_{0}}}%
\right\} ,  \label{L}
\end{equation}
where $C_{L}$ is independent from $u.$ By using (\ref{pp}) and
Bayes' rule
\begin{equation}
p_{1}\left( u|\ln S_{1}+h_{1}\right) =\frac{L\left( \ln
S_{1}+h_{1}|_{u}\right) p_{0}(u)}{\int L\left( \ln
S_{1}+h_{1}|_{u}\right) p_{0}(u)du}  \label{learn}
\end{equation}
(see \cite{Press}) one can find\textit{\ a }\textit{posteriori} pdf
of $u_{1} $ conditional on $\ln S_{1}+h_{1}$:
\begin{equation}
p_{_{1}}(u|\ln S_{1}+h_{1})=C_{1}\exp \left\{ -\frac{\left( \ln (\frac{S_{1}%
}{S_{0}})+h_{1}-\mu -\varphi h_{0}-u\right) ^{2}}{\sigma ^{2}e^{h_{0}}}%
\right\} \exp \left\{ -\frac{u^{2}}{2\sigma _{0}^{2}}\right\} ,
\label{le}
\end{equation}
where $C_{1}$ is independent of $u.$ Equation (\ref{le}) gives the
learning procedure that can be used at each stage of the process to
revise the probability density function for $u_{n}.$ By using
(\ref{le}) we can find the recurrence relation for $p_{_{n}}(u):$
\begin{equation}
p_{_{n+1}}(u)=C_{n+1}\exp \left\{ -\frac{\left( \ln (\frac{S_{n+1}}{S_{n}}%
)+h_{n+1}-\mu -\varphi h_{n}-u\right) ^{2}}{\sigma
^{2}e^{h_{n}}}\right\} p_{_{n}}(u).
\end{equation}
It is a well known property of Gaussian distribution (see
\cite{Press}) that this learning procedure gives a revised
probability density function that is also Gaussian:
\begin{equation}
p_{_{n}}(u)=\frac{d}{du}\mathbf{P}\left\{ u_{n}<u\right\} =\frac{1}{\sqrt{%
2\pi \sigma _{n}^{2}}}\exp \left\{ -\frac{\left( u-m_{n}\right) ^{2}}{%
2\sigma _{n}^{2}}\right\} .  \label{Gauss}
\end{equation}
The standard deviation $\sigma _{n}$ and the mean $m_{n}$ at
successive stages are given by recurrence equations
\begin{gather}
\sigma _{n+1}^{2}=\frac{\sigma ^{2}}{e^{-h_{n}}\sigma _{n}^{2}+\sigma ^{2}}%
\sigma _{n}^{2},  \label{sigma} \\
m_{n+1}=\frac{\sigma ^{2}}{e^{-h_{n}}\sigma _{n}^{2}+\sigma ^{2}}m_{n}+\frac{%
\sigma _{n}^{2}[\ln (\frac{S_{n+1}}{S_{n}})-\mu +h_{n+1}-\varphi h_{n}]}{%
\sigma _{n}^{2}+\sigma ^{2}e^{h_{n}}}.  \label{mue}
\end{gather}
At each discrete time $n$, the uncertainty about the value of
$u_{n}$ is described by the probability density function $p_{n}(u)$
given by (\ref {Gauss}) which is completely specified by the mean
value $m_{n}$ and the
standard deviation $\sigma _{n}.$ These state variables are the \textit{%
sufficient statistics}, and their transformation from on stage to
the next is given by equations (\ref{sigma}) and (\ref{mue}).
Equation (\ref{sigma})
shows that at every stage 
\begin{equation*}
\sigma _{n+1}^{2}<\sigma _{n}^{2},
\end{equation*}
that is, the variance of $u_{n+1}$ is smaller than the variance of
$u_{n}$. In other words the uncertainty about the innovation $u_{n}$
is reduced at every stage $n.$ This is crucial for option pricing.
Let us note that although $\sigma _{n}^{2}\rightarrow 0$ as
$n\rightarrow \infty ,$ stochastic volatility does not disappear
overall. Now we are in a position to apply the adaptive procedure
(\ref{sigma}) for the pricing of an European call option.
\section{Adaptive stochastic optimization}
Assume that an investor sells a European call option with strike
price $X$ \ for $C_{0}$ and invests the money in a portfolio
containing $\Delta _{0\text{ }}$shares and $\theta _{0}$ bonds. The
investor is concerned with hedging this position. It is well known
that in incomplete markets a portfolio replicating the payoff of the
option ceases to exist. Therefore the investor tries to find a
trading strategy that reduces the risk of an option position to some
intrinsic value. The value of the portfolio $V_{n}$ at time $n$ is
given by
\begin{equation}
V_{n}=\Delta _{n}S_{n}+\theta _{n}B_{n},\;V_{0}=C_{0}.
\end{equation}
Using the self-financed trading strategy condition
\begin{equation*}
(\Delta _{n+1}-\Delta _{n})S_{n+1}+(\theta _{n+1}-\theta
_{n})B_{n+1}=0,
\end{equation*}
one can obtain an equation for $V_{n}$:
\begin{equation}
V_{n}=(1+r)V_{n-1}+\Delta _{n-1}(e^{\mu +\sigma \delta _{n-1}e^{\frac{h_{n-1}%
}{2}}}-1-r)S_{n-1}.  \label{port}
\end{equation}
Let us recall the theory of risk-minimization in option pricing that
was developed in \cite{Muller,Folmer,Schweizer} (see also
\cite{BS,AS,YP}). The investor's purpose is to choose a trading
strategy $\{\Delta _{0},..,\Delta _{N-1}\}$ such that the terminal
value of the portfolio, $V_{N},$ should be as close as possible to
the options payoff: $(S_{N}-X,0)^{+}.$ Thus, the expected value of
their difference, under the ''real-world'' probability measure, must
be equal to zero: $\mathbf{E}\{(S_{N}-X,0)^{+}-V_{N}\}=0,$ while the
variance
\begin{equation}
R=\mathbf{E}\{((S_{N}-X,0)^{+}-V_{N})^{2}\}  \label{risk}
\end{equation}
as a measure of the risk should be minimized.
Let us consider the problem of minimizing the risk function $R$ for an $N$%
-stage process, starting from the initial states
\begin{equation}
S_{0}=S,\ \ \ \ \ V_{0}=V,\ \ \ \ \ h_{0}=h  \label{in}
\end{equation}
with \textit{a priori} probability density $p_{0}(u)$ specified by the mean $%
m_{0}=0$, and the standard deviation $\sigma _{0}=\sigma _{u}$. Here
we use a stochastic programming procedure proposed in \cite{Fed}.
Let us introduce the minimal risk
\begin{equation}
R_{N}(S,V,h,\sigma _{u})=\underset{\Delta _{0,.....},\Delta _{N-1}}{\min }%
\mathbf{E}\{((S_{N}-X,0)^{+}-V_{N})^{2}\}
\end{equation}
that can be achieved by starting from the initial state (\ref{in})
with a priori pdf $p_{0}(u)$. After the first decision $\Delta
_{0}=\Delta $ of the $N$-stage process we have
\begin{gather}
S_{1}=Se^{\mu +\sigma \delta e^{h/2}}, \\
V_{1}=(1+r)V+\Delta (e^{\mu +\sigma \delta e^{h/2}}-1-r)S, \\
h_{1}=\varphi h+u, \\
\sigma _{1}^{2}=\frac{\sigma ^{2}\sigma _{u}^{2}}{e^{-h}\sigma
_{u}^{2}+\sigma ^{2}}.
\end{gather}
The principal of optimality yields the general functional recurrence
equation (Bellman's equation)
\begin{eqnarray}
R_{N}(S,V,h,\sigma _{u}) &=&\underset{\Delta }{\min }\mathbf{E}%
\Bigl\lbrace%
%
R_{N-1}(Se^{\mu +\sigma \delta e^{h/2}},  \label{Bellman} \\
&&(1+r)V+\Delta (e^{\mu +\sigma \delta e^{h/2}}-1-r)S,\varphi h+u,\frac{%
\sigma ^{2}\sigma _{u}^{2}}{e^{-h}\sigma _{u}^{2}+\sigma ^{2}})%
\Bigl\rbrace%
%
.  \notag
\end{eqnarray}
(see \cite{Bellman,Ber,J}). By using the explicit expressions for
$\varphi (\delta )$ and $p_{0}(u)$, the equation (\ref{Bellman}) can
be rewritten as follows
\begin{multline}
R_{N}(S,V,h,\sigma _{u})=\underset{\Delta }{\min }%
\Bigl\lbrace%
%
\frac{1}{2\pi \sigma _{u}}\int_{-\infty }^{\infty }\int_{-\infty
}^{\infty
}R_{N-1}(Se^{\mu +\sigma \delta e^{h/2}}, \\
(1+r)V+\Delta (e^{\mu +\sigma \delta e^{h/2}}-1-r)S,\varphi
h+u,\frac{\sigma
^{2}\sigma _{u}^{2}}{e^{-h}\sigma _{u}^{2}+\sigma ^{2}})e^{-\frac{\delta ^{2}%
}{2}-\frac{u^{2}}{2\sigma _{u}^{2}}}d\delta du%
\Bigl\rbrace%
%
.  \label{E22}
\end{multline}
To solve (\ref{E22}), we need to know the value of the risk function $%
R_{N}(S,V,h,\sigma _{u})$ for $N=1.$ It follows from (\ref{risk})
that
\begin{equation}
R_{1}(S,V,h)=\underset{\Delta }{\min }\mathbf{E}\left\{ ((Se^{\mu
+\sigma \delta e^{h/2}}-X,0)^{+}-(1+r)V-\Delta (e^{\mu +\sigma
\delta e^{h/2}}-1-r)S)^{2}\right\} .  \label{one-stage}
\end{equation}
Using (\ref{ff}) we have
\begin{multline}
R_{1}(S,V,h)=\underset{\Delta }{\min }%
\Bigl\lbrace%
%
\frac{1}{\sqrt{2\pi }}\int_{-\infty }^{\infty }((Se^{\mu +\sigma
\delta
e^{h/2}}-X\text{ },0)^{+} \\
-(1+r)V-\Delta (e^{\mu +\sigma \delta
e^{h/2}}-1-r)S)^{2}e^{-\frac{\delta ^{2}}{2}}d\delta
\Bigl\rbrace%
%
,  \label{E23}
\end{multline}
where
\begin{equation}
(Se^{\mu +\sigma \delta e^{h/2}}-X,0)^{+}=%
\Bigl\lbrace%
%
_{0\text{ \ \ \ \ \ \ \ \ \ \ \ \ \ \ \ \ \ \ \ \ \ \ \ \ \ \ \ \ \
\ \ \ \ otherwise}}^{Se^{\mu +\sigma \delta e^{h/2}}-X\text{ \ \ \
for \ \ \ }\sigma ^{-1}e^{-h/2}\left( \ln (XS^{-1})-\mu \right)
<\delta }.
\end{equation}
The integral in (\ref{E23}) can be evaluated exactly (see Appendix
A). This allows us to find the explicit expressions for the optimal
policy, $\Delta _{1}(S,V,h)$ and the risk $R_{1}(S,V,h)$ when there
is one stage-to-go (see Appendix A). Now putting $N=2$ in equation
(\ref{E22}) and using the expression for $R_{1} $, one can find the
risk function $R_{2}:$
\begin{multline}
R_{2}(S,V,h)=\underset{\Delta }{\min }%
\Bigl\lbrace%
%
\frac{1}{2\pi \sigma _{u}}\int_{-\infty }^{\infty }\int_{-\infty
}^{\infty
}R_{1}(Se^{\mu +\sigma \delta e^{h/2}},  \notag \\
(1+r)V+\Delta (e^{\mu +\sigma \delta e^{h/2}}-1-r)S,\varphi h+u)e^{-\frac{%
\delta ^{2}}{2}-\frac{u^{2}}{2\sigma _{u}^{2}}}d\delta du%
\Bigl\rbrace%
%
\end{multline}
Note that at each stage, the risk function $R_{n}$ does not depend
on the mean $m_{n}$. What is more, adaptation procedure starts only
at the third
step, when the risk $R_{3}$ becomes a function of $\sigma _{u}$ that is $%
R_{3}=R_{3}(S,V,h,\sigma _{u})$. The procedure can be repeated any
number of times to give the solution of the problem for any value of
$N$. The attractive feature of this algorithm is the simplicity with
which the adaptation procedure can be applied. The initial
investment $V$ determining a fair option price $C=V$ can be obtained
from the equation
\begin{equation}
\frac{\partial R_{N}\mathbf{(}S,V,h,\sigma _{u})}{\partial V}=0.
\end{equation}
In particular, for a one stage process ($N=1$), after minimization,
we obtain
\begin{multline}
V(S,h)=\frac{1}{2\sqrt{\pi }(1+r)}e^{-\mu -\frac{e^{h}\sigma ^{2}}{2}}%
\Bigl(%
%
\sqrt{2}e^{^{\mu +\frac{3e^{h}\sigma ^{2}}{2}}}+e^{e^{\mu
+\frac{e^{h}\sigma
^{2}}{2}}}\sqrt{\pi }(\ln (\frac{X}{S})+ \\
S(1-2r+2\mu )+(\mu +e^{h}\sigma ^{2}-\ln (\frac{X}{S}))(2\mathcal{N}(d)-1)%
\Bigr)%
%
,
\end{multline}
where $\mathcal{N}(d)$ is the cumulative distribution function for a
Gaussian variable:
\begin{equation*}
\mathcal{N}(d)=\frac{1}{\sqrt{2\pi }}\int_{-\infty }^{d}e^{-\frac{s^{2}}{2}%
}ds,\;\;d=\frac{(\mu +e^{h}\sigma ^{2}-\ln
(\frac{X}{S}))e^{-h/2}}{\sigma }.
\end{equation*}
The above results can be compared to those corresponding to the
standard model without an adaptive procedure. In the later case, an
\textit{a priori} density function for $u$ (\ref{pp}) is kept at
each stage, and the risk function $R_{N}$ becomes the function of
$S,V,$ and $h$ only. The Bellman recurrence equation for the risk
minimization problem is then given by
\begin{multline}
R_{N}(S,V,h)=\underset{\Delta }{\min }%
\Bigl\lbrace%
%
\frac{1}{2\pi \sigma _{u}}\int_{-\infty }^{\infty }\int_{-\infty
}^{\infty
}R_{N-1}(Se^{\mu +\sigma \delta e^{h/2}}, \\
(1+r)V+\Delta (e^{\mu +\sigma \delta e^{h/2}}-1-r)S,\varphi h+u)e^{-\frac{%
\delta ^{2}}{2}-\frac{u^{2}}{2\sigma _{u}^{2}}}d\delta du%
\Bigl\rbrace%
%
,  \label{E28}
\end{multline}
where $R_{1}$ is the same as (\ref{E23}). To illustrate our adaptive
control method we value the European call option with the strike
price $X=50$, the initial log-volatility value $h=0.1$, the interest
rate $r=0.05$, the expected return $\mu =0.1$, the volatility
parameter $\sigma =0.2$, the maturity of the option $T=1$, and $\varphi =0.1$%
. We also calculate the option price for the constant volatility case $(h=0)$%
. Figure 1 shows the results for the option price as a function of
$S$ for different number of steps of the adaptive (learning)
procedure. To illustrate the usefulness the adaptive approach, we
computed the value of a European call option for the standard
(no-learning) procedure using equation (\ref{E28}). In Figure 2 we
show the difference between the option prices with and without
adaptation. The number of steps $N=12.$ It is clear that the
adaptive procedure leads to a decrease in option price. Let us now
discuss the ``smile" effect arising from the above methodology, that
is, implied volatility varies with strike price. Using equations
(24) and (29) for $N=12$, we can retrieve various call option
prices, $C_{adap}(X)$, for different strike prices $X$. Implied
volatility $I=\sigma(X)$, can then be computed by inverting the
formula
\begin{equation}
C_{BS}(X,I)=C_{adap}(X),
\end{equation}
where $C_{BS}$ is the usual Black-Scholes price. We plot our results
in Figure 3, where we get a ``smile" curve, with a minimum near
at-the-money. Thus, the adaptive methodology presented here is shown
to reproduce the ``smile" effect observed in the market. There is a
one-to-one relationship between the volatility ``smile" and the
implied distribution. The volatility ``smile" in Figure 3 implies
some kurtosis and skewness in the implied distribution.
\section{Conclusions}
In contrast to most stochastic volatility models we applied an
adaptive control procedure which allows us to revise the stochastic
characteristics of latent volatility during decision making. We
assumed that the statistical properties of \ an innovation term in a
log-volatility equation are not known initially, but instead we have
an a priori estimation for them. By using Bayesian analysis, we
derived the recurrence equation for the variance of innovation term.
This equation describes a reduction of uncertainty about volatility
which is crucial for option pricing. We implemented the idea of
adaptive procedure by using the risk-minimization analysis and
stochastic dynamic programming. We showed that the adaptation leads
to a decrease in the option price compared to the standard models
without learning. The adaptive algorithm allows the investor to
hedge his position in a consistent way between two extremes: a
completely uncertain volatility and an ideal situation of known
volatility. Of course this paper leaves many open questions how to
model the uncertainties in a stochastic volatility setting. For
example, one can introduce the uncertainty in the  stochastic return
(\ref{aspr}) rather than in the latent volatility (\ref{vol}). We
hope to address these questions in the future works.
\pagebreak \renewcommand{\theequation}{A-\arabic{equation}} %
\setcounter{equation}{0}
\section*{Appendix A}
To evaluate the integral in (\ref{E23}), we need to split it into
two integrals. That is,
\begin{multline}
R_{1}(S,V,h)=\underset{\Delta }{\min }%
\Bigl\lbrace%
%
\frac{1}{\sqrt{2\pi }}\int_{-\infty }^{\sigma ^{-1}e^{-h/2}\left(
\ln (XS^{-1})-\mu \right) }(-(1+r)V-\Delta (e^{\mu +\sigma \delta
e^{h/2}}-1-r)S)^{2}e^{-\frac{\delta ^{2}}{2}}d\delta + \\
\frac{1}{\sqrt{2\pi }}\int_{\sigma ^{-1}e^{-h/2}\left( \ln
(XS^{-1})-\mu \right) }^{\infty }\left( Se^{\mu +\sigma \delta
e^{h/2}}-X-(1+r)V-\Delta
(e^{\mu +\sigma \delta e^{h/2}}-1-r)S\right) ^{2} \\
e^{-\frac{\delta ^{2}}{2}}d\delta
\Bigl\rbrace%
%
.
\end{multline}
By using \textit{Mathematica }one can get the following expression for%
\textit{\ }$R_{1}:$
\begin{multline}
R_{1}(S,V,h)=\underset{\Delta }{\min }\{\frac{e^{-l}}{2\sqrt{2\pi }}%
\Bigl(%
%
2e^{\mu +\frac{3e^{h}\sigma ^{2}}{2}}S(1+\Delta )(-2(1+r)V+\ln (\frac{X}{S}%
)\Delta + \\
S(1+\mu +\Delta (e^{h}\sigma ^{2}-2r+\mu )))\sigma +((V+\ln (\frac{X}{S}%
)+S(-1+r\Delta -(1+\Delta )\mu ))^{2}+e^{h}S(1+\Delta )^{2}\sigma ^{2})%
\mathcal{N}(d)%
\Bigr)%
%
+ \\
\frac{e^{l}}{2\sqrt{2\pi }}%
\Bigl(%
%
(%
\Bigl(%
%
-2e^{\mu +\frac{3e^{h}\sigma ^{2}}{2}}S\Delta (-2(1+r)V+\Delta (\ln (\frac{X%
}{S})+S(-1+\mu )))\sigma + \\
\sqrt{2\pi }(((1+r)V+S\Delta (\mu +e^{h}\sigma
^{2}))^{2}+e^{h}S^{2}\Delta
^{2}\sigma ^{2})(2-2\mathcal{N}(d)%
\Bigr)%
%
\Bigr)%
%
\end{multline}
where
\begin{equation*}
\mathcal{N}(d)=\frac{1}{\sqrt{2\pi }}\int_{-\infty }^{d}e^{-\frac{s^{2}}{2}%
}ds,\;\;d=\frac{(\mu +e^{h}\sigma ^{2}-\ln (\frac{X}{S}))e^{-h/2}}{\sigma},%
\text{ \ }l=\mu +\frac{e^{h}\sigma ^{2}}{2}
\end{equation*}
($\mathcal{N}(d)$ is the cumulative distribution function for the
normal distribution). Differentiation with respect to $\Delta $
leads to the optimal first decision when there is one stage-to-go,
$\Delta _{1}(S,V,h)$, starting from the initial state $S$ and $V$:
\begin{multline}
\Delta _{1}(S,V)=%
\Bigl(%
%
e^{-l}%
\Bigl(%
%
2S%
\Bigl(%
%
e^{\mu +\frac{3e^{h}\sigma ^{2}}{2}}S(r-\mu )\sigma e^{h/2}
-e^{l}\sqrt{2\pi
}((r-\mu )((1+\mu)V+\ln (\frac{X}{S})-\mu +e^{h}\sigma ^{2}) \\
e^{h}S\sigma ^{2})%
\Bigr)%
%
+2e^{l}\sqrt{2\pi }S((r-\mu )(S-\ln (\frac{X}{S})+S\mu )-e^{h}S\sigma ^{2})(2%
\mathcal{N}(d)-1)%
\Bigr)%
%
\Bigr)%
%
\\
(\sqrt{2\pi }S^{2}((r^{2}+e^{h}\sigma ^{2}))^{-1},  \notag
\end{multline}
and substituting this in the expression for $R_{1}(S,V,h)$ gives
\begin{multline}
R_{1}(S,V,h)=\frac{e^{-l}}{2\sqrt{2\pi }}%
\Bigl(%
%
2e^{\mu+\frac{e^{h}\sigma^{2}}{2}}S((1+r)V-\ln(\frac{X}{S})+S(1+2\mu
)\sigma
+ \\
e^{l}\sqrt{2\pi })(V+rV+\ln(\frac{X}{S})+S(-1+r+\mu
))^{2}+e^{h}S^{2}(1+r)^{2}\sigma ^{2}+  \notag \\
e^{l}\sqrt{2\pi }%
\Bigl(%
%
((1+r)V+S(r-\mu ))^{2}+e^{h}S^{2}\sigma ^{2}+((S-\ln(\frac{X}{S}))  \notag \\
(-2(1+r)V-\ln(\frac{X}{S})+e^{h/2}S(1-2r+2\mu ))+e^{h}S^{2}(1+\sigma ^{2})2%
\mathcal{N}(d)%
\Bigr)%
%
\Bigr)%
%
.  \notag
\end{multline}
\pagebreak

\pagebreak \linespread{1.0}
\begin{figure}[!t]
\centering
\includegraphics[scale=1.3]{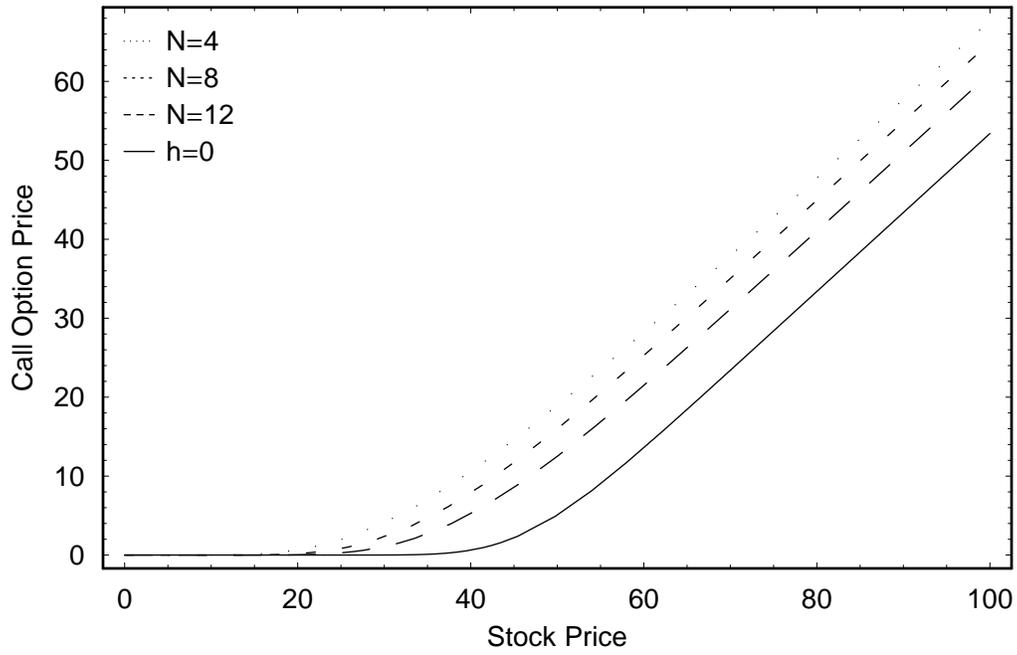}
\caption{ Option price as a function of asset price for different
number of stages of the adaptive process.} \label{sub-fig-test}
\end{figure}
\linespread{1.0}
\begin{figure}[!t]
\centering
\includegraphics[scale=1.3]{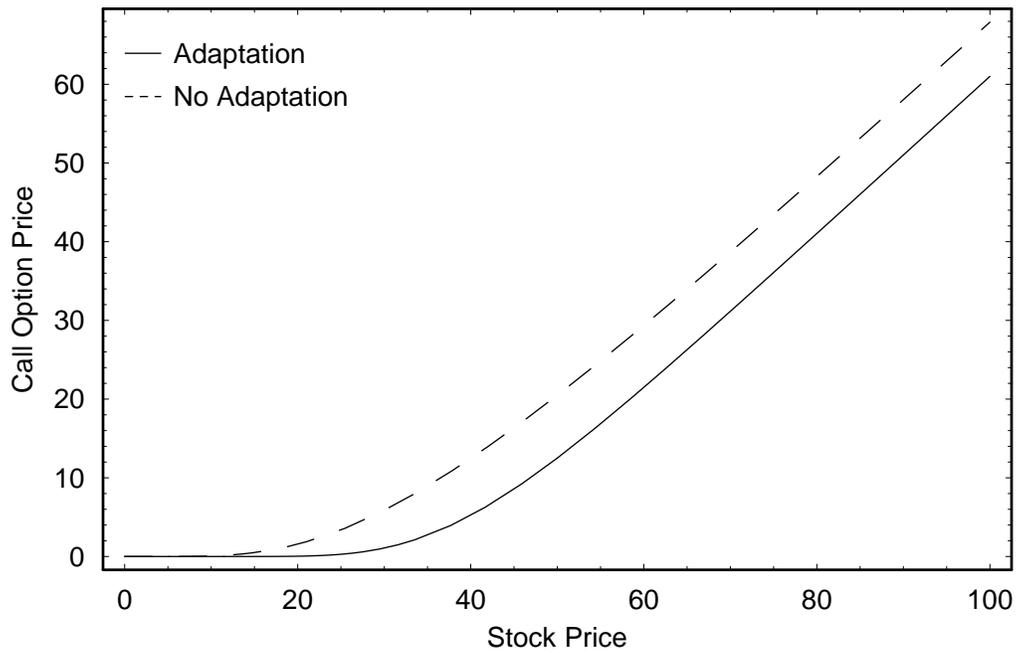}
\caption{Comparison of option price with and without adaptation for
$N=12$.}
\end{figure}
\begin{figure}[t]
\centering
\includegraphics[scale=1.3]{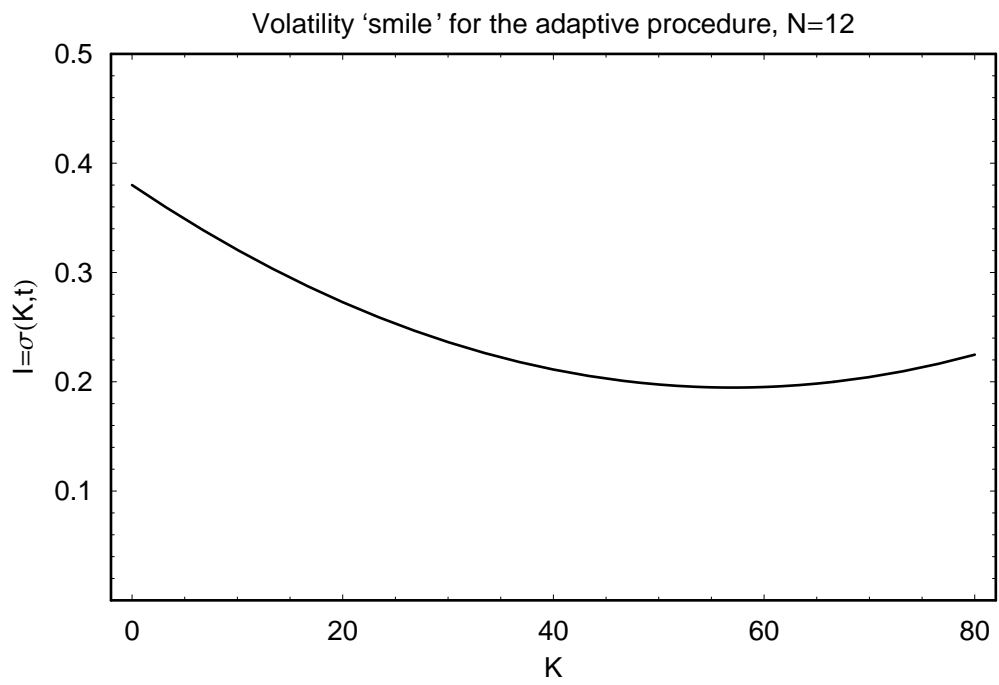}
\caption{Implied volatility $I=\sigma(X)$ as a function of strike
price $X$. The constants are taken as in Figures 1 and 2. }
\end{figure}
\end{document}